\documentclass[twocolumn,pre]{revtex4}
\usepackage{amsmath,amssymb}
\usepackage{graphicx,pstricks}
\usepackage{url}
\usepackage{bm}
\include{epsf}

\def\(({\left(}
\def\)){\right)}                       
\def\[[{\left[}
\def\]]{\right]}

\renewcommand{\>}{\rangle}
\newcommand{\beq}{\begin{equation}}
\newcommand{\eeq}{\end{equation}}
\newcommand{\bea}{\begin{eqnarray}}
\newcommand{\eea}{\end{eqnarray}}

\begin{document}
\title{Quantifying lymphocyte receptor diversity}

\author{Thierry Mora$^{1}$, Aleksandra M. Walczak$^{2}$
}

\affiliation{$^{1}$ Laboratoire de physique statistique,
    CNRS, UPMC and \'Ecole normale sup\'erieure, 24, rue Lhomond,
    Paris, France}
\affiliation{$^{2}$ Laboratoire de physique th\'eorique,
    CNRS, UPMC and \'Ecole normale sup\'erieure, 24, rue Lhomond,
    Paris, France}

\begin{abstract}
 % abstract
To recognize pathogens, B and T lymphocytes are endowed with a wide repertoire
of receptors generated stochastically by V(D)J
recombination. Measuring and estimating the diversity of these
receptors is of great importance for understanding adaptive immunity. In this
chapter we review recent modeling approaches for analyzing receptor
diversity from high-throughput sequencing data. We first
clarify the various existing notions of diversity, with its
many competing mathematical indices, and the different biological
levels at which it can be evaluated.
We then describe
inference methods for characterizing the statistical diversity of
receptors at different stages of their history: generation,
selection and somatic evolution. We discuss the intrinsic difficulty
of estimating the diversity of receptors
realized in a given individual from incomplete samples. Finally, we emphasize the limitations of 
diversity defined at the level of receptor sequences, and advocate the more relevant notion
of functional diversity relative to the set of recognized
antigens.

\end{abstract}

\maketitle

 % maintext

\section{Introduction}
To protect its host against pathogens, the adaptive immune system of jawed vertebrates expresses a large repertoire of distinct receptors on its B- and T lymphocytes. These receptors must recognize a wide range of pathogens to trigger the response of the adaptive immune system. Since each receptor is specialized in recognizing specific pathogens, a very diverse repertoire of receptors is required to cover all possible threats. While one can now sequence the repertoires of individuals with some depth, it remains unclear how to quantify or even define their diversity, and what aspects of this diversity are relevant for recognition. These fundamental questions are further obscured by the purely technical but important issue of reliably sampling immune repertoires. 

The actual number of lymphocytes varies from species to species, but in all cases is large. Estimates of the number of T cells in humans are of the order of $3\cdot 10^{11}$ cells \cite{Jenkins2010}. Each cell expresses only one type of receptor. 
Cells proliferate and form clones, so that many distinct cells may share a common receptor. As we will discuss further, the number of unique distinct receptors is very hard to estimate. However, even a conservative lower bound of $10^6$ unique receptors \cite{Robins2009,Warren2011} is much larger than the total number of genes in the human genome ($\sim 20,000$).
This broad diversity of receptors is not hard-coded, but is instead generated by a unique gene rearrangement process that couples a combinatoric choice of genomic templates with additional randomness.

\begin{figure}
\noindent\includegraphics[width=1\linewidth]{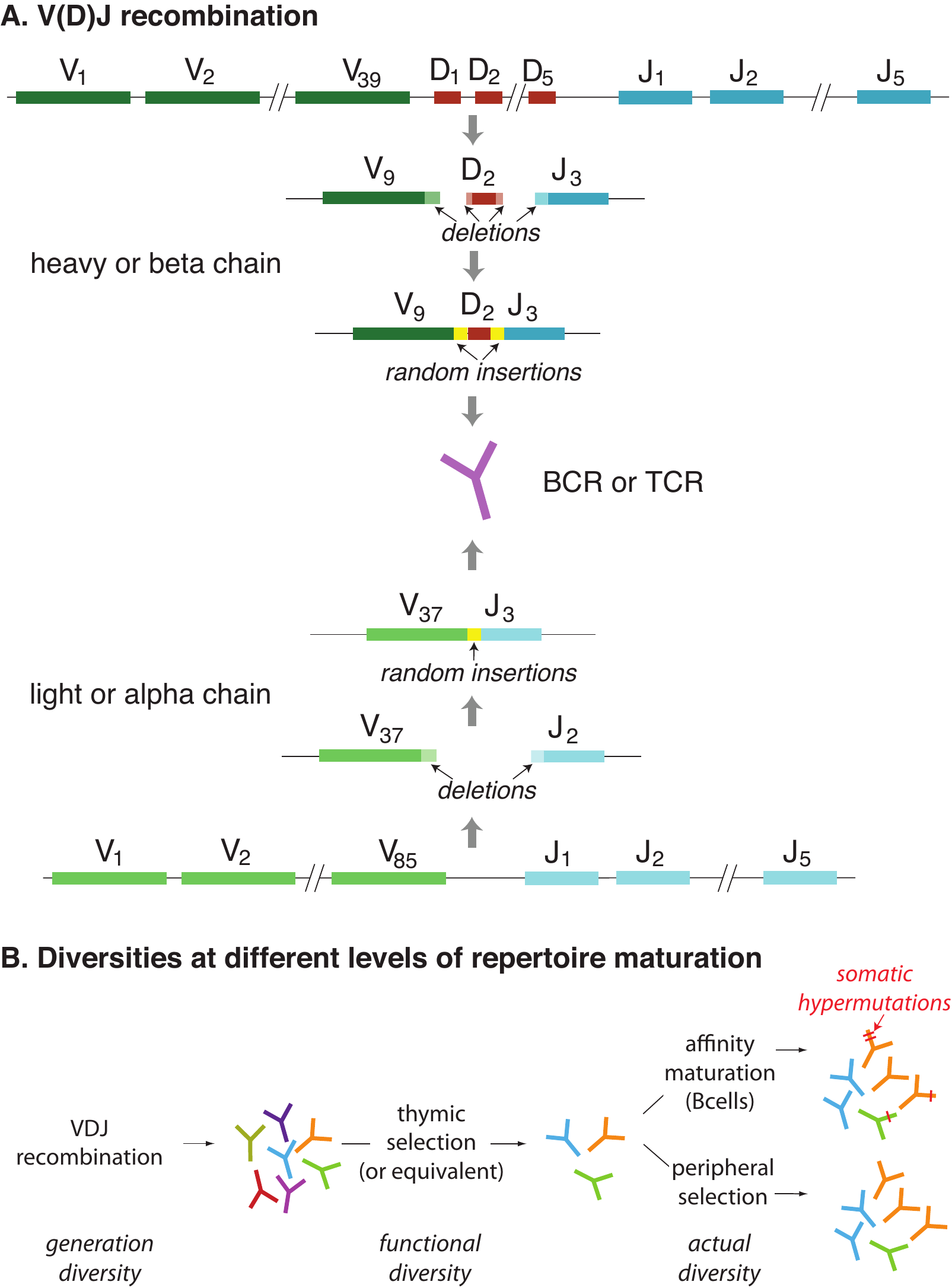}
\caption{{A. V(D)J recombination of T- and B cell receptors (TCR and BCR).} TCRs and BCRs are made of two chains, one shorter and one longer, called the $\alpha$ and $\beta$ chains for TCRs, and the light and heavy chains for BCRs. Each chain is obtained by a gene rearrangement process called V(D)J recombination, by which two (for the shorter chain) or three (for the longer chain) segments are assembled together from palettes of templates encoded in the genome. At each of the junction between these segments, further diversity is added by stochastic deletions and insertions of random, non-templated nucleotides.
{ B. Evolution of repertoires of TCR and BCR.} After their generation by V(D)J recombination, receptors first pass a selection process, called thymic selection for TCRs, whereby nonfunctional and self-reactive receptors are discarded. They are then released into the periphery, where they may divide, die, proliferate and differentiate as a function of the signals they receive from antigens or other immune cells. In addition, BCRs are subject to somatic hypermutations as B cells mature in germinal centers following an infection.}
\label{fig1}
\end{figure}

Each receptor is made up of two arms:  B-cell receptors (BCR) have a light and a heavy chains, while  T-cell receptors  (TCR) have analogous $\alpha$ and $\beta$ chains. Each chain is composed of three segments called V, D and J in the case of heavy or $\beta$ chains, and two segments V and J in the case of light or $\alpha$ chains.
These segments are combinatorically picked out of several genomic templates for each type, in a process called V(D)J recombination \cite{Hozumi1976}, as schematized in Fig.~\ref{fig1}A.
This recombination is achieved by looping DNA and excising the template genes that lie between the selected gene segments. In the case of heavy or $\beta$ chains, the D-J junction is assembled first, followed by the V-D junction. The precise number of templates for each segment differs from species to species, but generally results in a combinatoric diversity of $\sim 1000$ for each chain. This combinatoric assortment is followed by stochastic nucleotide deletions and insertions at the junctions between the newly assorted V-D and D-J fragments (or V-J fragment for the shorter chain), forming what is termed junctional diversity. This stochastic step largely increases the repertoire diversity, as we will show in detail. As a result of this procedure the receptor DNA may be out-of-frame, or the encoded protein may not be functional or correctly folded. The newly assembled $\beta$ chain sequences then are tested with a surrogate $\alpha$ chain for their binding and expression properties. If they pass this selection step, the second chain is assembled and the whole receptor undergoes a similar round of selection against proteins that are natural to the organism, or self proteins. Receptors that do not bind any self-protein or bind too strongly to self-proteins are discarded. If a receptor fails these tests, the cell may attempt to recombine its second chromosome.

The processes of recombination and selection are stochastic, and therefore are characterized by their own intrinsic diversity, which we may view as a statistical or potential diversity. It is distinct from the diversity realized in a given individual at a given time, with its finite number of recombined receptors, much like the potential diversity of the English language is distinct from -- and much larger than -- the diversity of texts found in a single library. While most previous discussions, with the expection of \cite{Zarnitsyna2013}, have focused on the realized rather than potential diversity of receptors, in this chapter we will discuss both.

After generation and selection, B- and T cells feed the naive repertoire where they attempt to recognize foreign antigens (Fig.~\ref{fig1}B). The dynamics of lymphocytes vary widely between B and T cells, as well as between species. However, a common feature is that cells whose receptors successfully bind to antigens proliferate, producing either identical offspring (T-cells) or that differ by somatic point hypermutations (B-cells). A fraction of the cells that have undergone proliferation are kept in what is called the memory repertoire,
while cells that have not received a proliferation signal stay in the naive repertoire. Cells that share a common receptor, or ``clonotype,'' define a clone. The clonal structure of the lymphocyte repertoire is one of the characteristics of repertoire diversity.

The diversity of lymphocyte receptors can be studied with the help of repertoire high-throughput sequencing experiments \cite{Weinstein2009,Robins2009,Freeman:2009fja,Robins2010}, which have been developing rapidly over the last few years \cite{Benichou2012,Warren2013a,Six2013,Woodsworth2013b,Georgiou2014a,Calis2014}. 
These experiments focus on the region of the chain that encompasses the junctions between the recombined segments, allowing for the complete identification of the receptor chain. This region includes the Complementarity Determining Region 3 (CDR3), defined from roughly the end of the V segment to the beginning of the J segment, which is believed to play an important role in recognition.
Because sequence reads can only cover one of the two chains making up the receptor, most studies have focused on the diversity of one chain at a time. However, new techniques make it possible to pair the two chains together \cite{Dekosky2014,Turchaninova2013c,Howie2015}, opening the way for the analysis of repertoires of complete receptors.
In general, a tissue (blood, lymph node, thymus, germinal center, etc.) sample is taken and the mRNA or DNA of the lymphocytes of interest are sorted out. 
Different technologies have been developed for DNA and mRNA. Data are usually clustered and error-corrected for PCR and sequencing errors \cite{Shugay2014}.
Many recent experiments use unique molecular barcodes associated to each initial mRNA molecule, which help correct for PCR amplification noise \cite{Vollmers2013,Egorov2015,Best2015c}, and allow for the direct measurement of relative clone sizes using sequence counts. Unless an error occurred in the first round of PCR, barcodes can reliably pick up even very rare sequences, as long as they are present in the sample. These experiments result in a list of unique receptor chain sequences, and if the data was barcoded, of reliable counts for the corresponding number of RNA molecules in the initial sample. This information is the staring point for the analysis of repertoire diversity.

In this chapter we discuss approaches for estimating repertoire diversity from the datasets generated by these new technologies. We first review and discuss the different definitions of diversity -- species richness, entropy, and other diversity indices -- and their relation to the distribution of clonotype frequencies. We also emphasize the need to distinguish the different levels at which diversity may be evaluated: recombination diversity, post-selection potential diversity, actual diversity realized in a particular individual, in a particular tissue, or with a particular phenotype, etc. 
We review recent efforts to  calculate accurately the  diversity of receptors generated by V(D)J recombination using high-throughput sequencing data. We discuss the challenges of estimating diversity when the clonal structure is scale-free, as is generically the case in many reported cases. We conclude by discussing the importance of sequence diversity and contrast it with more biologically relevant but elusive notion of functional diversity.

\section{A family of diversity measures}
A number of different diversity measures have been proposed to quantify the vastness of lymphocyte repertoires \cite{Greiff2015a,Yaari2015,Greiff2015b}: the Shannon entropy \cite{Mora2010}, the Simpson index \cite{davenport-2007}, and most commonly the total number of clonotypes or species richness \cite{Arstila1999,Robins2009,Warren2011,Qi2014,Laydon2014}. These diversity measures are taken from ecology, where they are used to quantify the diversity of species. They are all related to a generalized family of diversity measures called the R\'enyi entropy \cite{Renyi1961}, parametrized by $\beta$ and defined as:
\beq
H_\beta=\frac{1}{1-\beta}\ln\left[\sum_s p(s)^{\beta}\right],
\eeq
where $p(s)$ is the probability, frequency or abundance of a given receptor sequence or clonotype $s$. For $\beta\to 1$ we recover Shannon's entropy:
\beq
H_1=-\sum_s p(s)\ln p(s).
\eeq
The exponential of the R\'enyi entropy defines a generalized class of diversity indices called Hill diversities \cite{Hill1973}:
\beq
D_\beta=\exp[H_\beta].
\eeq
This index can be interpreted as an effective number of clonotypes in the data. For $\beta=1$, it is simply the exponential of Shannon's entropy, and we will refer to it as Shannon's diversity.
For $\beta=2$, it reduces to the inverse of Simpson's diversity index, $D_2=1/\sum_s p(s)^2$. The Simpson index gives the probability that two sequences drawn at random from the distribution are identical, and is related to a common measure of inequality, the Gini-Simpson index, defined as $1-1/D_2$.
$D_0$ is the species richness, while $D_{\infty}=1/\max_s p(s)$ is the inverse of the Berger-Parker index.

Each of these diversity indices is a summary statistics of the information contained in the distribution of clonotype frequencies, {\em i.e.} the distribution of values of $p(s)$ themselves. 
This frequency distribution may in fact be viewed as the most complete description of the diversity of the repertoire.
Conversely, the whole spectrum of R\'enyi entropies $H_\beta$ is sufficient to reconstruct the full clonotype frequency distribution. In other words, the functions $H_\beta$, $D_\beta$, and the distribution of frequencies carry the exact same information \cite{Mora2016}.
The choice of a single diversity measure $D_\beta$, rather than the full frequency distribution, is often useful to make comparisons between individuals, tissues, experiments, etc. When $\beta$ is large enough, it may also be less sensitive to experimental noise than the frequency distribution. 

\begin{figure}
\noindent\includegraphics[width=1\linewidth]{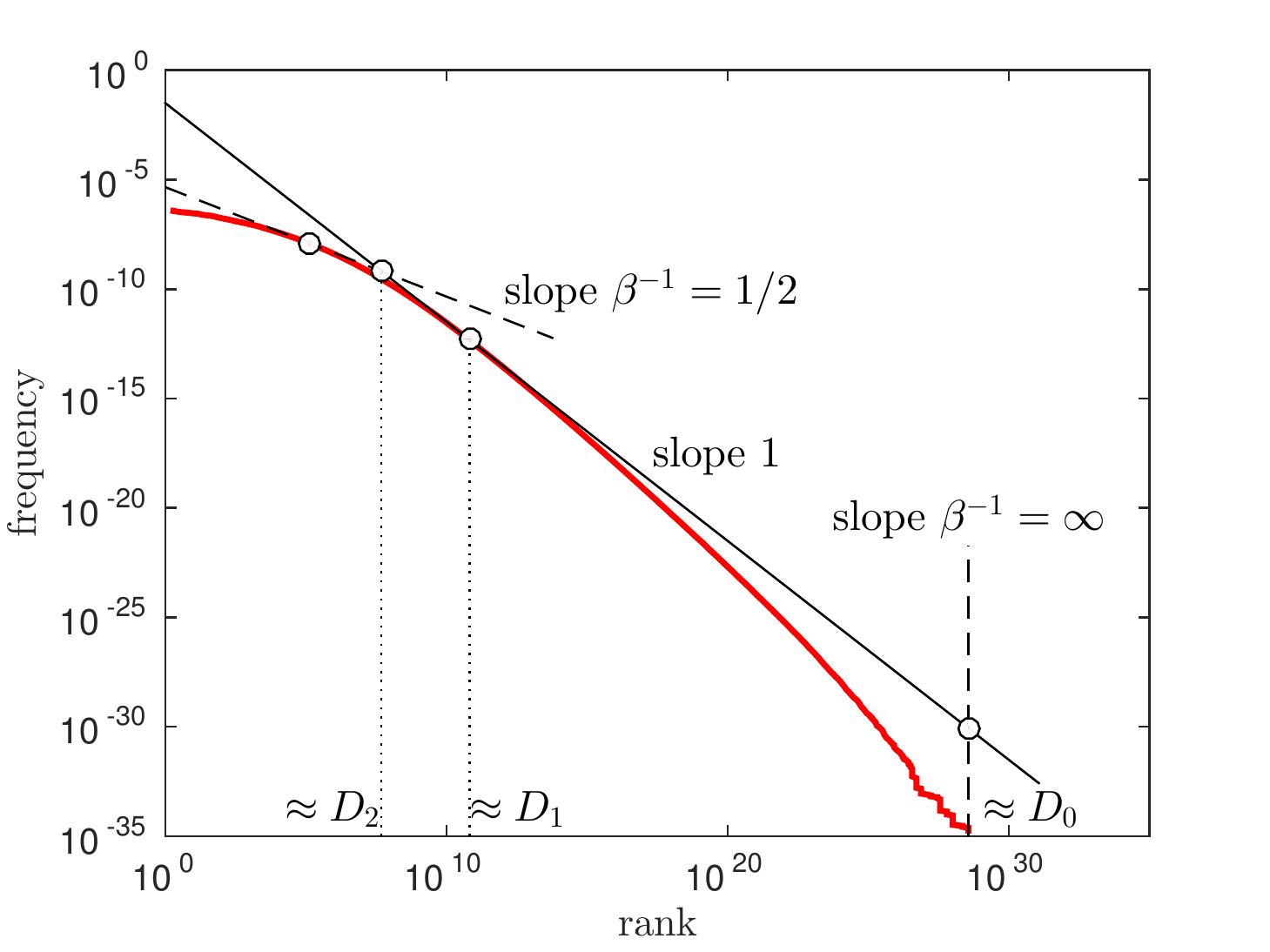}
\caption{Geometric construction of Hill diversities from a rank-frequency curve. The Hill diversity of order $\beta$, $D_\beta=[\sum_s p(s)^\beta]^{1/(1-\beta)}$, can be approximated from the intersection between the tangents of slope -1 and $-1/\beta$. $D_0$ is the total number of types or species richness; $D_1$ is the exponential of Shannon's entropy; and $D_2$ is the inverse of Simpson's diversity index.}
\label{fig2}
\end{figure}

It is possible to get a rough estimate of Hill diversities by simple inspection of the frequency distribution, represented as a rank-frequency graph with a double logarithmic scale \cite{Mora2016}. A simple geometric construction, illustrated by Fig.~\ref{fig2}, helps understand the meaning of the various indices, what properties of the underlying cumulative clone size distribution they are most likely to capture, and where one should stop trusting them because of insufficient sampling.
The intersection of the the tangents of slope $-1$ and $-\beta^{-1}$  to the rank-frequency curve gives the Hill diversity index $D_\beta$. This construction emphasizes the fact that different diversity measures focus on sequences of various frequencies: large values of $\beta$ tend to favor very common clonotypes, while low values favor rare ones. Geometrically, tangents of small slopes (large $\beta$, e.g. Simpson's index or Shannon's entropy) osculate the rank-frequency curve at high frequencies, while large slopes do so at low frequencies.
Thus, diversity indices $D_\beta$ with a small $\beta$ rely very strongly on correclty capturing the tail of rare clonotypes. This is particularly true for $D_0$, the species richness, which is very hard to estimate as it requires to estimate the number of unseen clonotypes. This observation warns us against the pitfalls of estimating diversity when dealing with incomplete samples. The larger the $\beta$, the more reliable the Hill index $D_\beta$ should be. In general, estimates of the species richness $D_0$ should be taken with extreme caution, as we will further discuss in concrete examples.

\section{Quantifying V(D)J recombination}
The repertoire is a dynamic ensemble of receptors that evolves somatically. As the repertoire is shaped, its diversity changes significantly. Repertoires at different functional stages, from generation to memory, show different levels of potential and realized diversity. By analyzing unique receptors from high-throughput sequencing data, one can track these changes. We start by decribing the diversity of the initial stochastic recombination of receptors.

Each cell has two sets of chromosomes. If the first V(D)J rearrangement results in a non-functional receptor, the second one recombines \cite{Janeway}. When this second rearrangement is successful, the cell expresses the functional receptor, but keeps the rearranged nonfunctional DNA. This nonfunctional receptor is expressed at a basal, leaky level despite allelic exclusion, especially for $\alpha$ chains, and may also be captured by  genomic DNA sequencing. 
These out-of-frame receptors offer unique insight into the raw generation process, because they were
 never selected for, as they owe their survival to the gene expressed from the other chromosome.
We can therefore use these sequences to gain insight into the generation process, and analyze the {\em potential} diversity of recombination, {\em i.e.} the statistics of unique receptors that can ever be formed as a result of V(D)J recombination. As already noted, this diversity of the generation process should not be confused with the actually realized diversity in a given individual, which is generically smaller.

As the numbers will show, the recombination probability of each generated sequence is so small that it is hopeless to sample their distribution by simply counting how often we observe them. Besides, this counting number is not expected to reflect the frequency of generation alone, because of lymphocyte population dynamics. As we pointed out, cell proliferation is independent of the identity of the out-of-frame sequence of interest, and in the limit of infinite data should not in principle affect such an estimate. However, for any dataset coming from a single individual, these heterogeneities in the clone size completely dominate the sequence counts. For this reason, it is suitable to count each unique sequence only once to remove these possible biases. Starting with a dataset of unique realizations of the recombination process, we need a model to describe their probability distribution. This model is based on what we know about the recombination process: choice of V(D)J segments, stochastic number of deletions of each gene segments, stochastic number and identities of inserted nucleotides at each junction. 
Thus, taking the simpler case of $\alpha$ or light chains, the probability of a given recombination scenario $r$ can be written as:
\beq
P_\text{rearr}(r) = P(V,J) P(\text{del}V|V) P(\text{del}J|J) P({\rm ins}),
\label{model_eq}
\eeq
where $\text{del}V$ and $\text{del}J$ denote the number of deletions at the V and J ends, and ``${\rm ins}$'' is the list of inserted nucleotides. A very similar expression accounting for three genes and two junctions can be written for the $\beta$ or heavy chains. The form of the model is motivated by biophysical considerations: the number of deletions of the $J$ end does not depend on the choice of the V segment, the number and identities of insertions does not depend on the gene choice, and follows a Markov chain. These assumptions, however, should and can be checked consistently by verifying that no correlations in the data remains unaccounted for by the model \cite{Murugan2012}.

The parameters of the generation model \eqref{model_eq} cannot be directly read off the sequences, because it is impossible in general to assign with certainty a recombination scenario to a given sequence, as many distinct scenarios can lead to the same sequence through convergent recombination \cite{Venturi:2006hk}. As we will quantify below, this effect is very significant and cannot be ignored. Importantly, it forces us to think of scenarios or sequence annotation in a probabilistic manner, rather than try to select the most probable one as is often done in annotation software \cite{Volpe2006a,Gaeta2007,Munshaw2010}. The generation parameters can be inferred using a standard implementation of the Expectation-Maximization algorithm, an iterative procedure that maximizes the likelihood of the data. The algorithm works by
collecting summary statistics about the elements of the recombination scenarios to build the model distribution \eqref{model_eq}. The recombination scenarios are themselves assigned probabilistically using the previous iteration of the model. The algorithm, which relies on the enumeration of all plausible scenarios giving rise to each sequence, is computationnally heavy, but can be significantly sped up after mapping the problem onto a hidden Markov model and using standard dynamic programming tools \cite{Elhanati2016}.

Once a recombination model such as Eq.~\ref{model_eq} has been inferred, it can be used to generate and analyse sequences with the same statistical properties as the original data. It can also be used to quantity the various types of diversity indices discussed in the previous section. Note that, because of convergent recombination,
the diversity of generated sequences is expected to be smaller than the diversity of the scenarios that produce them. The generation probability of a sequence $s$ is given by the sum of the probabilities of all scenarios that could have given rise to this sequence:
\beq
P_{\rm gen}(s)=\sum_{r\to s} P_{\rm rearr}(r).
\eeq
The diversity measures calculated from $P_{\rm gen}$ and $P_{\rm rearr}$ are therefore distinct.

\begin{figure*}
\noindent\includegraphics[width=.7\linewidth]{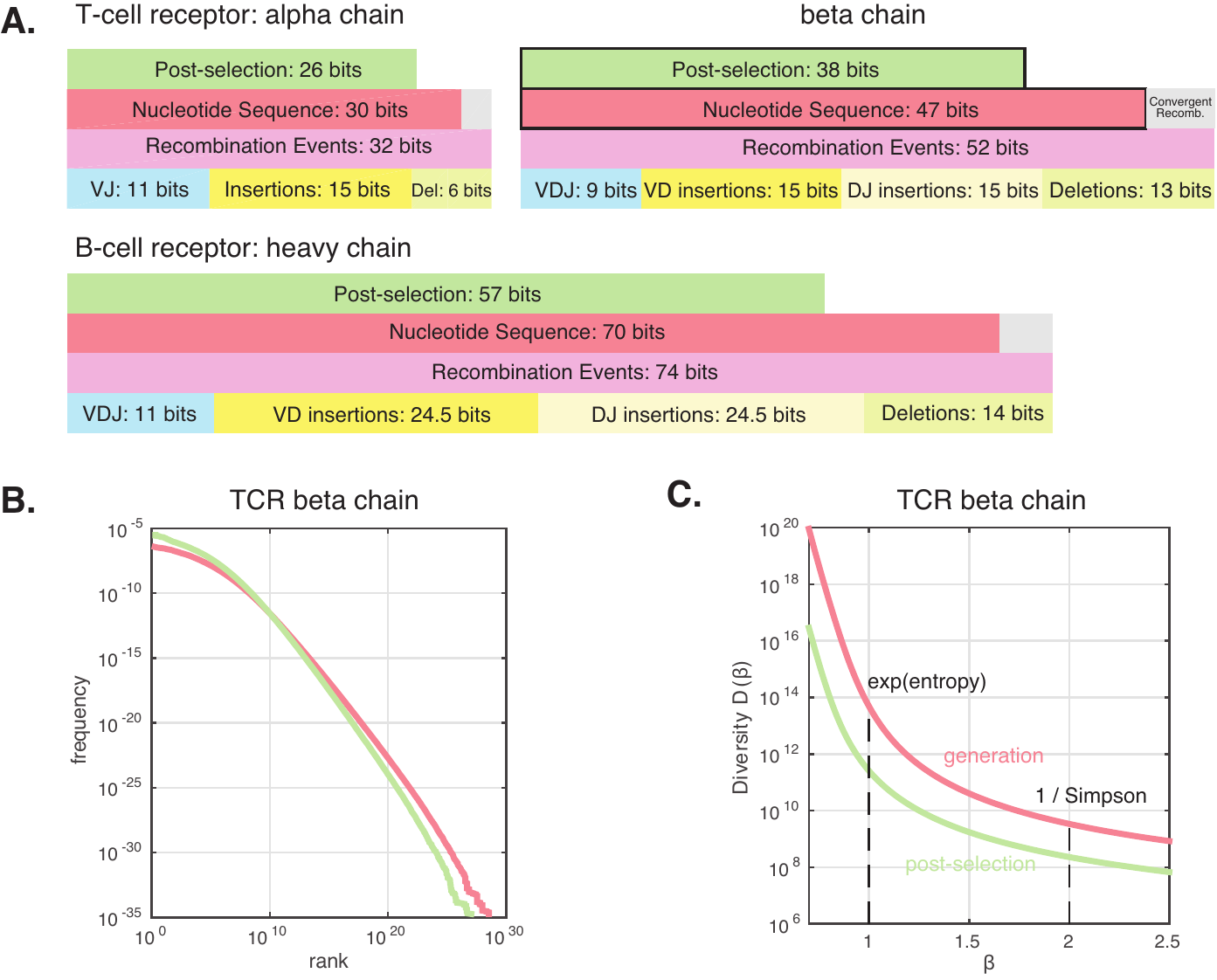}
\caption{Entropies and diversity indices of the receptor generation and selection process. A. Entropy of the V(D)J recombination process in TCR $\alpha$ and $\beta$ chains, and in BCR heavy chains. The entropy of recombination events (purple) can be decomposed into contributions for the choice of the V(D)J genes (blue), the number and identity of insertions (yellow), and deletions (light green). The sequence entropy (red) is slightly smaller than the recombination entropy because several recombination events can lead to the same sequence (convergent recombination, in gray). Following thymic selection, or the B-cell counterpart, the entropy is further reduced (green). B. Rank-frequency curves of TCR $\beta$ chain sequences, upon generation (red), and following thymic selection (green). C. Hill diversities for the same statistical ensembles. The Shannon diversity $D(1)$ is the exponential of the entropies shown in black boxes in A.
}
\label{fig3}
\end{figure*}

Recombination models have been inferred for T cell $\beta$ \cite{Murugan2012} and $\alpha$ \cite{Elhanati2016} chains, as well as for B cell heavy chains \cite{Elhanati2015}. In all these cases, the distributions inferred from different individuals were found to be surprisingly similar, with some variability in the gene segment usage, but very reproducible deletion and insertion profiles, consistent with a common biophysical mechanism of enzyme function. The entropy $H_1$ of sequences and recombination scenarios obtained from these models are reported in Fig.~\ref{fig3}A. Because the distribution of scenarios \eqref{model_eq} is a product of its various elements (gene choice, deletions, insertions), its entropy can also be broken up into their respective contributions. The entropy difference between recombination events (in purple) and sequences (in red), is the entropy of convergent recombination (in gray), which quantifies the diversity of scenarios resulting in the same sequence. For example, it is 5 bits for TCR $\beta$ chains, corresponding to a fairly large Shannon diversity number, $D_1\sim 30$. Note that the total number of possible scenarios for a given sequence, $D_0$ is much larger, but its precise definition depends on the cutoff we impose on the possible number of deletions and insertions.

Diversity in the heavy chain of B-cells is larger than that of T-cells. This difference can be attributed to longer CDR3 regions due to many more insertions at the junctions between the genes. The receptor generation process is characterized by an entropy of $\sim 70$ bits for  BCR heavy chains and $\sim 43$ bits for TCR $\beta$ chains. These numbers correspond to a Shannon diversity index  $D_1 \sim 10^{21}$ and $\sim 10^{14}$, respectively.

Although most studies have focused on the Shannon diversity index $D_1$, the full diversity spectrum of the generation process can be calculated. In  Fig.~\ref{fig3}B we show the rank-frequency curve of human TCR $\beta$ chains,
taken from Ref.~\cite{Mora2016} based on the model of Ref.~\cite{Murugan2012}. As explained in the previous section, the full range of diversity indices $D_\beta$ can be calculated from that curve, and are shown in Fig.~\ref{fig3}C. In addition to the Shannon diversity $D_1$ already discussed, of special interest is the inverse of the Simpson index, $D_2$. The Simpson index corresponds to the probability that the same nucleotide sequence is obtained from two independent draws. It gives the expected number of shared sequences between two individuals, normalized by the product of their repertoire sizes, assuming that their receptor sequences were generated independently from the same source. Thus, it is deeply linked to the notion of ``public'' sequences found in several individuals, and making up the public repertoire \cite{Venturi:2006hk,davenport-2007,Venturi:2011co}.
This number, estimated to be $1/D_2\sim 3\cdot 10^{-10}$ for human TCR $\beta$ chains from the model, is in fact very close to that measured in the data for out-of-frame sequences \cite{Murugan2012}.

It is important to stress that, however large, these numbers are {\em not} the total number of possible receptor sequences, $D_0$, which is much larger. As we can see from the rank-frequency plot of generated TCR $\beta$ chain sequences (Fig.~\ref{fig3}B, red), generation probabilities span over 20 orders of magnitude. The largest rank of $\sim 10^{30}$ is in fact a lower bound to $D_0$ limited by the finite sampling of sequences by the model. To better estimate $D_0$, one may count the total number of possible deletion profiles reported for each gene, and multiply that number by the total number of possible insertion profiles of at most $L_{\rm max}$ nucleotides, $(4^{L_{\rm max}-1})/3$, for each of the two junctions. Doing so with $L_{\rm max}=26$, the largest number of insertions reported in \cite{Murugan2012}, yields an upper bound of $D_0\sim 2\cdot 10^{39}$ for the TCR $\beta$ chain alone. However, because this estimate is very sensitive to the value of $L_{\rm max}$, which is not precisely known and may depend on the sample size, it must be taken with some caution.

The above estimates only include heavy or $\beta$ chains. Coupling this chain with the light or $\alpha$ chain adds further diversity.  Since the shorter ($\alpha$ and light) chains have only one junctional region between the V and J genes, their diversity is much lower. For example, TCR $\alpha$ chains were estimated to have a generation Shannon entropy of $H_1=30$ bits, or $D_1\sim 10^9$ \cite{Elhanati2016}. The part of the entropy that is attributable to the gene choice is similar to that reported for the $\beta$ chain, of the order of $10$ bits. While that contribution was only a small fraction of the overall diversity  for the $\beta$ chain, it is comparable to that of insertions  for the $\alpha$ chain. The number of possible $\alpha$ chain sequences can be estimated similarly to the $\beta$ chain, yielding $D_0\sim 5\cdot 10^{21}$. 

Assuming that the two chain rearrangements are independent, the overall diversity of the pool from which TCRs are generated is about $H_1\sim 75$ bits, or $D_1\sim 10^{23}$, and a total potential repertoire of size $D_0\sim 10^{61}$. Note that this last estimate is much larger than the classically quoted number of $10^{15}$ from \cite{Davis1988}, which assumed a much more restricted junctional diversity. Analysis of recently published $\alpha$-$\beta$ sequence pairings should allow for more precise estimates of these diversity numbers for TCRs \cite{Howie2015} and BCRs \cite{Dekosky2014}.

All these diversity numbers are very large. Clearly, a single individual is only able to sample a tiny fraction of the potential pool of receptor sequences, with a total T-cells count of $\sim 3\cdot 10^{11}$ in humans \cite{Jenkins2010}.

\section{Thymic selection and hypermutations}
After sequences have been generated by V(D)J recombination, they undergo an initial selection process. For T-cells, this takes place in the thymus and is called thymic selection. An analogous process occurs for B-cells. Sequences that bind too strongly to the host's own self-proteins, as well as those that bind too weakly to them, are discarded. By analyzing the in-frame naive receptor repertoire, one can study how the diversity of the repertoire is affected by this initial selection process. 
While the recombination diversity, $P_{\rm gen}(s)$, described the potential variability from the gene rearrangement process, this post-selection naive diversity, $P_{\rm sel}(s)$, describes the statistics of sequences actually found in the naive repertoire. It is still a potential diversity, as it refers to a statistical ensemble of receptors, rather than a finite set of receptors found in a given individual.

One can define a sequence-dependent selection factor $Q(s)=P_{\rm sel}(s)/P_{\rm gen}(s)$ quantifying how the distribution of sequences is affected by thymic selection. As before, sampling from $P_{\rm sel}(s)$ is impossible in practice because of the too large number of sequences, and models of the selection factor $Q(s)$ are needed. For example, it may take the factorized form
\beq\label{Q}
Q(s)=\prod_{i=1}^L q_{i;L}(a_i),
\eeq
where $(a_1,a_2,\ldots,a_L)$ is the amino-acid sequence of the CDR3 region of length $L$, and the single-position factors $q_{i;L}(a)$ are inferred from the data using maximum likelihood. This model describes very well the statistics of naive and memory TCR $\beta$-chain sequences \cite{Elhanati2014}, $\alpha$-chain sequences \cite{Pogorelyy2016}, and naive BCR heavy chain sequences \cite{Elhanati2015}. The selection factors $Q(s)$ were shown to depend only on the amino-acid rather than nucleotide sequence, consistent with our hypothesis that selection acts on the protein product and its functional properties (folding, stability, binding, etc.). Although selection factors may vary significantly from individual to individual in the statistical sense, these differences are relatively small. In addition, models inferred from the memory and naive sequence repertoires were found to be similar, suggesting that the selection factors $Q(s)$ capture universal functional properties of the receptor proteins.

Diversity numbers can be estimated from the model of Eq.~\ref{Q}. The entropy of the post-selection distributions of receptor sequences, $P_{\rm sel}(s)=Q(s)P_{\rm gen}(s)$ are shown in green in Fig.~\ref{fig3}A. The rank-frequency distribution and Hill diversities $D_\beta$ of the post-selection ensemble of TCR $\beta$ chain sequences are shown in green in Fig.~\ref{fig3}B and C.

Diversity is reduced by selection from $47$ to $38$ bits for TCR $\beta$ chains, from $30$ to $26$ bits for $\alpha$ chains, and from $70$ to $58$ bits for BCR heavy chains, corresponding to $D_1\sim 3\cdot 10^{11}$ for $\beta$ chains, $D_1\sim 7\cdot 10^{7}$ for $\alpha$ chains (or a combined TCR diversity of $2\cdot 10^{19}$ assuming independence between the two chains), and $D_1\sim 3\cdot 10^{17}$ for heavy chains. About 2 bits of this reduction are due to the removal of visibly nonfunctional sequences (out-of-frame or having stop codons). However, most of the diversity loss is caused by negative selection against sequences that were unlikely to be produced in the first place. Frequent sequences are enriched by the selection process, while rare ones are more likely to be removed. This enhancement of inequalities between sequences is the main source of entropy reduction by selection.

It should be noted that these estimate rely on an effective model \eqref{Q}, which may miss many important aspects of the selection process. In particular, negative selection, which prunes the repertoire of specific sequences that bind to self-antigens, is likely not accounted for by the model. This further diversity loss would be specific to each individual and its set of self-antigens, which depends on its HLA types. 
To assess whether all the aspects of selection that are not individual specific are well captured by Eq.~\ref{Q}, one can ask whether the Simpson index calculated with the model, $1/D_2$, is consistent with the observed repertoire overlap between distinct individuals,
 as it should if the two repertoires were drawn independently from the same distribution $P_{\rm sel}(s)$. Indeed the model and data showed good agreement  \cite{Elhanati2014}, confirming that the model describes the statistics of sequences accurately.

Following their release into the periphery, cells undergo a somatic evolution process by which they divide, die or proliferate depending on the signals they receive. In the case of T cells, it is not clear how this evolution affects the potential naive diversity, as TCR $\beta$-chain sequences expressed by memory cells are statistically indistinguishable from naive ones \cite{Elhanati2014}. In contrast, BCRs experience somatic hypermutations as B cells proliferate upon antigen recognition, during the process of affinity maturation. These hypermutations are stochastic but do not occur uniformly across the receptor, favoring instead sequence context dependent `hotspots' \cite{Shapiro1999,Cowell2000}. High-throughput repertoire sequencing now makes it possible to build predictive statistical models of hypermutations, by disentangling mutation from substitution rates using either synonymous mutants \cite{Yaari2013a} or out-of-frame sequences \cite{Dunn-Walters01031998,Elhanati2015}. 
Out-of-frame sequences have a raw mutation rate ranging from a 5\% to 10\%, implying an additional $0.4$ bits per nucleotide. This additional diversity is a huge boost if this estimate holds for the whole length of the receptor sequence.
However, the increase in diversity due to hypermutations should depend on how long cells have been allowed to evolve. As affinity maturation consists of alternating cycles of mutation and selection, the effects of hypermutations on diversity cannot entirely be decoupled from selective pressures. The inference of selection during affinity maturation using repertoire sequencing is currently a very active field of study \cite{Uduman01072011,Yaari2012a,Kepler14,Laserson2014,Uduman2014,Mccoy2015,Yaari2015,Yaari2015a}. 

\section{Realized diversity}
Thus far we have focused on the potential diversity of lymphocyte receptors. Its object is the probability that each receptor sequence has been generated, selected and, in the case of BCR, hypermutated into its final form. One can also study the realized diversity of receptor clonotypes actually present in a given individual at a given time.
The relative frequency of clonotypes in an individual can vary greatly depending on the history of cell divisions and deaths, and is in general distinct from the probabilities $P_{\rm gen}$ and $P_{\rm sel}$ discussed so far.
Measuring accurate clonotype frequencies relies on trustworthy counts made possible by unique molecular barcodes associated to original mRNA molecule \cite{Vollmers2013,Egorov2015,Best2015c} (with the caveat that cells may express variable amounts of mRNA molecules).
One can build the rank-frequency relation as before, by ranking clonotypes in a given individual from most common to rarest. This relation can be measured for different phenotypes (naive or memory, CD4 or CD8), in different tissues or organs, or at different ages, to study the organisation and evolution of diversity.

\begin{figure}
\noindent\includegraphics[width=1\linewidth]{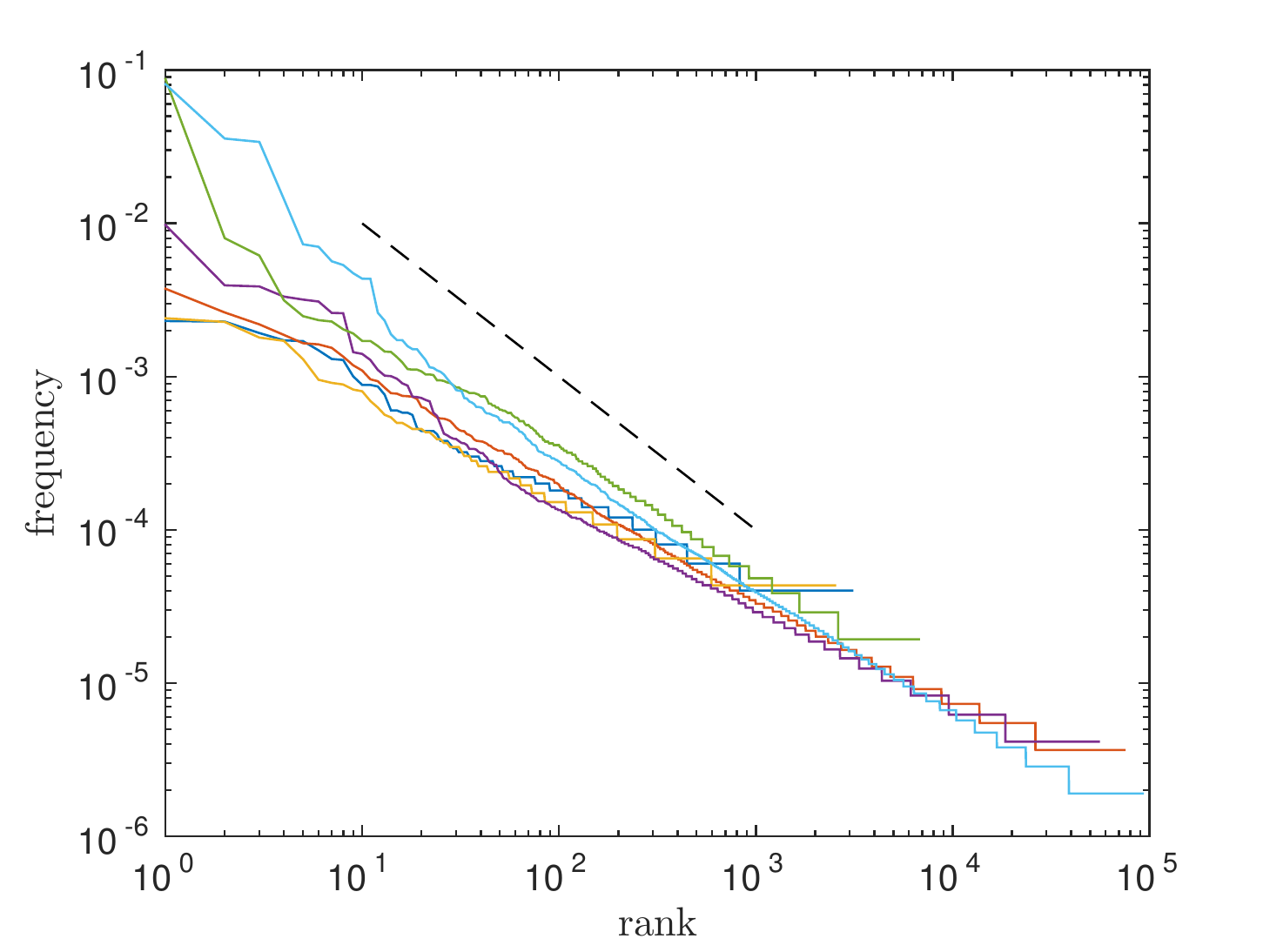}
\caption{Clonotype frequency vs. rank in the sequenced unpartitioned repertoires of six individuals from \cite{Pogorelyy2016}. These relations are close to a power law with exponents ranging from $-0.65$ to $-1$. The dashed line shows a slope of $-1$.}
\label{fig4}
\end{figure}

In Fig.~\ref{fig4} we plot the rank-frequency relation for the unpartitioned TCR $\beta$-chain repertoires sampled from the blood of six individuals \cite{Pogorelyy2016} and sequenced using unique molecular barcodes. A striking feature of these relations is that they seem to follow a power law, $f\propto 1/r^\alpha$, where $f$ and $r$ denote the clonotype frequency and rank, with exponent $\alpha$ ranging from $0.65$ to $1$, with a mean of $0.78$. This observation is consistent with previous reports on zebrafish BCR \cite{Weinstein2009,Mora2010} or mouse TCR repertoires \cite{Zarnitsyna2013}. 
These power laws cannot be explained by a neutral model in which cells divide and die stochastically at a constant rate. Instead, they are consistent with models where each clone evolves under a fluctuating fitness shaped by its changing antigenic environment \cite{Desponds2016}.

Power-law frequency distributions make it challenging to estimate diversity measures $D_\beta$ \cite{Mora2016}. This difficulty can be understood by considering the geometric construction of diversities of Fig.~\ref{fig2}: examining the rank-frequency curve of Fig.~\ref{fig4}, no tangent of slope $-1$ can be easily defined. Mathematically, the normalization of the distribution strongly depends on the maximal rank, as $\sum_r 1/r^\alpha$ is a diverging series, meaning that the distribution is dominated by a very large number of very small clonotypes. This is particularly problematic as these rare clonotypes are not well captured by incomplete sampling.

Most past studies of repertoire diversity have actually focused on the hardest diversity measure to estimate in the face of these sampling issues, namely the species richness index $D_0$. By sequencing a subset of the repertoire with low-throughput techniques and extrapolating to the entire repertoire, Arstila and collaborators found a lower bound to the total size of the TCR repertoire of $10^6$ distinct $\beta$ chains, each pairing to 25 distinct $\alpha$ chains, {\em i.e.} $2.5\cdot 10^7$ distinct TCRs \cite{Arstila1999}. This bound has since been revisited using high-throughput sequencing data, yielding the same order of magnitude of a few millions \cite{Robins2009,Warren2011}.

In practice, most experiments are performed on samples of blood or tissues and do not sequence every single cell. Even experiments using a whole tissue are subject to losses. 
The problem of species richness estimation from incomplete samples is not specific to lymphocyte repertoires and has been extensively discussed in ecology.
A number of estimators of $D_0$, such as Chao1 \cite{Chao1984}, the abundance-based coverage estimator \cite{Chao1992}, or more recently DivE proposed in the context of TCRs \cite{Laydon2014}, have been developed to address this issue. Another estimator using multiple samples, Chao2 \cite{Chao2002}, has recently been used to yield a lower bound of $10^8$ distinct TCR $\beta$ chains in humans \cite{Qi2014}.
All these estimators implictly assume that the distribution of frequencies is reasonably peaked, and may not be appropriate for broad distributions such as power laws.

To illustrate the inadequacy of most estimators to capture the true species richness of power-law distributed clone sizes, we numerically generated $D_0=10^7$ distinct clonotypes, and fixed their abundance to
\beq\label{powerlaw}
C_r=(D_0/r)^{\alpha},
\eeq
where $r=1,\ldots,D_0$ is the rank of the clonotype ordered by abundance, and $\alpha=0.8$ to mimick the data of Fig.~\ref{fig4}. We simulated a sample comprising $1\%$ of the entire dataset, by drawing $S_r$, the size of clonotype of rank $r$ in the sample, from a Poisson distribution of mean $C_r/100$. We calculated Chao1,
\beq
D_0\approx D^{\rm raw}_0+\frac{n_1^2}{2n_2},
\eeq
where $D^{\rm raw}_0$ is the number of sampled clonotypes ($S_r>0$, $n_1$ is the number of singletons ($S_r=1$), and $n_2$ the number of doubletons ($S_2=2$). This estimate gave $D_0=3\cdot 10^6$ instead of the true value of $10^7$. Dividing the dataset into 5 subsamples as in \cite{Qi2014}, and calculating Chao2  yields a similar estimate, $3.2\cdot 10^6$. The reason for this underestimation is deep and does not depend much on the details of the estimator. When downsampling, one loses information about the rare clones, which dominate the species richness. Extrapolating their number from larger clones must rely on implicit or explicit assumptions about the clonal distribution, which are likely not satisfied by fat-tailed distributions such as power laws.
It is therefore likely that most current estimates from high-throughput sequencing data are only lower bounds to the true species richness.

In fact, simple theoretical arguments based on thymic output estimates and neutral models of clonal evolution give upper bounds of $10^{10}$-$10^{11}$\cite{Kesmir2000,Lythe2015}.
However, since we have argued that the power-law in the rank-frequency curve did not support the hypothesis of neutrality, it is legitimate to ask what species richness would be predicted from a power-law distribution of clone sizes. Assuming that the rank-size relation is given by Eq.~\ref{powerlaw}, the average clonotype size reads:
\beq
\<C\>=\frac{1}{D_0}\sum_{r=1}^{D_0}{\left(\frac{D_0}{r}\right)}^{\alpha}\approx {D_0^{1-\alpha}}\int_{1}^{D_0} \frac{1}{r^\alpha} =\frac{1}{1-\alpha},
\eeq
where we have approximated the sum by an integral, which is valid for large $D_0$. Plugging $\alpha=0.8$ gives an average clone size of 5 cells, and hence a species richness $D_0=3\cdot 10^{11}/5\sim 10^{11}$ of the same order of magnitude as total number of T cells. Note however that this estimate is very sensitive to the value of $\alpha$, as the average clone size becomes $\sim\ln(D_0)$ for $\alpha= 1$, and $\sim\zeta(\alpha)D_0^{\alpha-1}$ for $\alpha>1$, where $\zeta(\alpha)$ is the Riemann zeta function.

Although the validity of the power law across the entire spectrum of clone sizes is a matter of debate, this example emphasizes the need for models to extrapolate the size distribution to the very rare clonotypes, the knowledge of which is essential for evaluating species richness.

\section{Towards a functional diversity}

All the diversities discussed in this chapter apply to nucleotide sequences. 
These estimates demonstrate the potential of the adaptive immune system to generate a huge diversity of sequences, while identifying the biases of their generation and selection. However, they do not directly inform us about the functional diversity of the repertoire, defined as its capacity to recognize a wide variety of antigens. First of all, the binding properties of receptors are determined by their amino-acid sequences, the diversity of which is smaller due to the degeneracy of the genetic code. But more fundamentally,
a given antigen can be recognized by many receptors --- a phenomenon termed cross-reactivity or polyspecificity. Mason \cite{Mason1998}  argued that if not for cross-reactivity, an individual would need a repertoire as large as the number of antigens it can encounter, or $\sim 10^{15}$ for TCRs, which is well beyond the number of lymphocytes a human or a mouse can afford. 
Simple models can help estimate the minimal size of the functional repertoire \cite{Perelson1979,DeBoer1993,Zarnitsyna2013}. Theoretical arguments also suggests
that cross-reactivity gives a certain freedom in the identity and binding properties of the receptors, implying that two individuals experiencing similar antigenic environments need not share common receptors through the convergent evolution of their repertoires \cite{Mayer2015}.

Quantifying the functional diversity of the repertoire is arduous because it requires to precisely characterize cross-reactivity by mapping the sequence of receptors to their binding properties. The identification of TCRs that bind to specific antigens using tetramer experiments in mouse \cite{Moon2007203} shows that a single antigen is bound by 20-200 out of $4\cdot 10^7$ CD4+ T cells, {\em i.e} a fraction $5\cdot 10^{-7}$-$5\cdot 10^{-6}$ of the total population. Conversely, a single TCR can recognize many antigens. A lower bound of $10^6$ has been reported for an autoimmune TCR from a human patient \cite{Wooldridge2012a}, but that number must be much larger ($>5\cdot 10^{-7}\times 10^{15}=5\cdot 10^8$) so that the TCR repertoire may cover the entire set of possible peptides. 

Assessing cross-reactivity in a more quantitative and systematic way requires to massively measure the binding properties of a huge numbers of receptor-antigens pairs.
High-throughput mutational scans combining binding assays with next-generation sequencing technologies 
now make it possible to measure the binding properties of a single receptor against many peptides \cite{Birnbaum2014}, or of many mutagenized receptors againt a single antigen \cite{Adams2016}.
Integrating these measurements into predicitve models of receptor-antigen binding would provide powerful tools for analysing lymphocyte repertoires. The diversity of receptor sequences could then be augmented by the more relevant diversity of antigens that can be recognized by them, with varying potencies and frequencies.

This work was supported in part by grant ERCStG n. 306312, and by the
National Science Foundation under Grant No. NSF PHY11-25915 through
the KITP where part of the work was done.

\bibliographystyle{pnas}

\end{document}